\documentstyle[12pt]{article}

\newcommand{\ba}{\begin{array}}
\newcommand{\ea}{\end{array}}
\newcommand{\be}{\begin{equation}}
\newcommand{\ee}{\end{equation}}
\newcommand{\bea}{\begin{eqnarray}}
\newcommand{\eea}{\end{eqnarray}}
\newcommand{\beas}{\begin{eqnarray*}}
\newcommand{\eeas}{\end{eqnarray*}}
\newcommand{\ra}{\rightarrow}
\newcommand{\ep}{\epsilon}

\font\blackboard=msbm10 at 12pt
\font\blackboards=msbm7
\font\blackboardss=msbm5
\newfam\black
\textfont\black=\blackboard
\scriptfont\black=\blackboards
\scriptscriptfont\black=\blackboardss

\begin{document}

\thispagestyle{empty}

\pagestyle{plain}
\setcounter{page}{1}

\baselineskip18pt

\begin{flushright}
QMW-PH-98-44\\
hep-th/9901008\\
\end{flushright}
\vspace{1cm}

\begin{center}
{\Large\bf  Orientifolds of Matrix theory and \\ Noncommutative Geometry}\\
\vspace{1cm}
{\large Nakwoo Kim
\footnote{email address: {\tt N.Kim@qmw.ac.uk}}
}\\
\vspace{0.5cm}
{\it Physics Department}
\\ {\it  Queen Mary and Westfield College }
\\ {\it Mile End Road}
\\ {\it London E1 4NS UK} \\
\end{center}
\vskip 1.5 cm
\begin{abstract}
We study explicit solutions for orientifolds of Matrix 
theory compactified on noncommutative torus.
As quotients of torus, cylinder, Klein bottle and M\"obius
strip are applicable as orientifolds. We calculate the solutions
using Connes, Douglas and Schwarz's projective module solution,
and investigate twisted gauge bundle on quotient spaces as well. 
They are Yang-Mills theory on noncommutative torus with proper
boundary conditions which define the geometry of the dual space.
\end{abstract}
\newpage
\setcounter{page}{1}
\renewcommand{\thefootnote}{\arabic{footnote}}
\setcounter{footnote}{0}

\section{Introduction}
According to the Matrix theory conjecture \cite{bfss}\cite{suss}, 
discrete lightcone quantization (DLCQ) of M-theory is described by the 
maximally supersymmetric gauge U($N$) quantum mechanics, where $N$
is the lightlike momentum, or the number of D-particles when
interpreted as effective dynamics of D0-branes.

Toroidal compactification of M-theory using Matrix theory formulation
can be performed by considering D0-brane dynamics on the covering
space and imposing periodicity on the variables \cite{taylor}. It 
is shown to lead to Yang-Mills gauge theory on dual torus \cite{taylor},
when we consider ${\bf T}^d$ with $d < 4$. Additional moduli from
winding mode of extended objects in M-theory should be taken into
account when we consider compactification on higher dimensional
tori \cite{highcomp}.
Also the supersymmetric Yang-Mills (SYM) theory on ${\bf T}^2$ is 
modified when the three-form potential of eleven dimensional 
supergravity is turned on along the lightlike direction. It is
described by SYM theory on noncommutative torus \cite{CDS,dhull}. 
Noncommutative torus ${\bf T}^2_\theta$ has additional SL(2,{\bf Z}) 
symmetry, which corresponds to the 
T-duality in the DLCQ direction. In general it is in mathematical language
Morita equvalence of noncommutative tori \cite{schwarz} 
which governs the duality
of Matrix theory compactifications with nonvanishing expectation
value of NS-NS two form potential.
When compactified on ${\bf T}^d$ the complete Morita
group is found to be SO($d,d|${\bf Z}) \cite{riesch}. 
The noncommutative ${\bf T}^d$ is defined by $d\times d$ matrix
$\Theta$, which transforms with fractional linear transformations with
respect to SO($d,d|{\bf Z}$). 
The rank of the gauge group, magnetic flux numbers and instanton
or other higher topological characters together comprise 
spinor representation, which means that under this duality SYM theories
with different rank of the gauge group are related. Various related topics
such as 
D-brane dynamics and noncommutative geometry, Morita equvalence of
noncommutative tori and the duality symmetry of Matrix theory action
and the BPS spectrum are studied further by various authors
\cite{howuwu,howu,casa,ckrogh,aashe,ho,ko,bigatti,szabo,mozu,
aashe2,bmz,bm,hofver,hofver2,konsch}.

In this paper we study orientifolds of Matrix theory compactification.
As the simplest but still very illuminating examples, which are
important on their own, we consider
orientifolds of SYM on noncommutative ${\bf T}^2_\theta$. Matrix theory 
compactification on non-orientable surfaces had been studied in \cite{zwart}
\cite{krnonori} before the noncommutative geometry nature was noted.
After that it was noted that when we introduce the concept of noncommutativity,
the dual space may not be determined uniquely \cite{howuwu}. When we mod out
the torus to make cylinder, we can have the dual space either cylinder
of Klein bottle. In the same way both cylinder and Klein bottle can
be assigned as the dual space of Klein bottle. Dual space of M\"obius strip
is always again M\"obius strip. This kind of ambiguity is 
generic with M-theory \cite{witm,dasm} and Matrix theory 
\cite{kimrey1,kimrey2,kimrey3}. 
We have to resort to physical arguments
to decide which is the right answer. Usually we have to introduce properly
chosen twisted sector to cancel the anomaly. For example we know that 
Matrix theory on cylinder is SYM on cylinder with twisted sector fermions
on the boundary. These correspond to the 
$E_8$ gauge field sector introduced on the end of the world when we 
consider M-theory compactified on ${\bf S}^1/{\bf Z}_2$.
Similarly the dual of Klein bottle cannot be
cylindrical, since it is generically anomalous without twisted sector
fields which M-theory compactification lacks. 

In this paper we aim to solve the orientifold compactification 
equations, first using the projective module solution presented 
in \cite{CDS}, and then we construct twisted gauge bundle on 
noncommutative torus and Klein bottle.

\section{Compactification of Matrix Theory}
The procedure of compactification in Matrix theory is straightforward.
It is the philosophy of the Matrix theory, or D-brane dynamics, that
the positions of the D-branes are encoded in the matrices as the 
eigenvalues. To realize periodicity we demand the following conditions.
We consider compactification on two-torus in this paper.
\bea
X_1 + 2\pi R_1 &=& U_1 X_1 U^{-1}_1, \nonumber \\
X_2 + 2\pi R_2 &=& U_2 X_2 U^{-1}_2. \label{comdef}
\eea
That is, we identify $X_i +2\pi R_i$ with $X_i$, up to a certain
similarity transformation. There are 8 other directions in Matrix
theory but they are intact with above similaraty transformations.
So we have as well
\begin{eqnarray*}
X_1 &=& U_2 X_1 U^{-1}_2, \\
X_2 &=& U_1 X_2 U^{-1}_1, \\
X_a &=& U_i X_a U^{-1}_i, \;\;\; i=1,2, \;\;\; a=3,...10.
\end{eqnarray*}
It is obvious that with matrices of finite size one cannot satisfy
above conditions. The original solution \cite{taylor} was based on 
the assumption that the translation operators commute
$$ [U_1, U_2]=0. $$
The standard solution is taking $U_i$ as the generators of the algebra
of the functions on dual torus ${\bf \tilde{T}}^2$,
$$ U_i = e^{2\pi i R_i x_i}. $$
The particular solutions to Eq.(\ref{comdef}) are partial derivatives,
and general solutions should be covariant derivatives
\bea
X_1 &=& i \partial_1 + A_1(U_1,U_2), \nonumber \\
X_2 &=& i \partial_2 + A_2(U_1,U_2). 
\eea
Looking at $U_i$, we can see that dual space with size $\tilde{R_i}
=1/R_i$ is created. While $X_1,X_2$ become the covariant derivatives on
the dual torus, other components become scalar fields on the dual torus.

In general we can consider the case when the translation operators
$U_i$ do not commute each other, but satisfy
\be U_1 U_2 = e^{2\pi i \theta} U_2 U_1. \label{ntc} \ee
The physical meaning of the parameter $\theta$ was first studied in
\cite{CDS}. It is the integral of the three-form potential of
eleven dimensional supergravity on a three-cycle including the
lightlike direction,
$$ \theta = R \int C_{12-} dx^1 dx^2 dx^- , $$
Interpreted as functions on torus again, i.e. $U_i=e^{i\sigma_i}$, 
Eq.(\ref{ntc}) defines a quantum plane algebra,
$$ [\sigma_1, \sigma_2] = -2\pi i \theta . $$
Then this gives SYM theory on noncommutative torus, 
where the multiplication is defined as 
\be
fg \rightarrow \exp \left. \left( \pi i \theta \epsilon^{ij}
\frac{\partial}{\partial x_i ' }
\frac{\partial}{\partial x_j ''}
\right)
f(x ') g(x '') \right|_{x'=x''=x} .
\ee
\section{CDS' projective module solution}  
We start by reviewing the compactification solution for 
$e^{2\pi i \theta} \ne 1$, presented by Connes, Douglas and 
Schwarz (CDS) in \cite{CDS}. After we fix $U_i$, the general solution
has the form of $X_i = \bar{X_i} + A_i$, where $\bar{X_i}$ are 
particular solutions and $A_i$ are operators commuting with
$U_i$. We consider operators on the space of functions 
on ${\bf R} \otimes {\bf Z}_q$, where ${\bf Z}_q = {\bf Z} / q{\bf Z}$.
We define $U_i$ as operators
acting on function $f(s,k)$ where $s \in {\bf R},
k \in {\bf Z}_q$ and transforming them as 
\bea
U_1 f(s,k) &=& e^{2\pi i \gamma s} f(s,k-p), \nonumber \\
U_2 f(s,k) &=& e^{-2\pi i k/q} f(s+1,k) . 
\eea
They satisfy 
\be
U_1 U_2  = e^{-2\pi i \gamma + 2\pi i p/q} U_2 U_1 ,
\ee
and we set $\theta = p/q - \gamma$.
We define the operators $\bar{X}_i$ as follows 
\bea 
\bar{X}_1 f(s,k) &=& i \nu \frac{\partial f}{\partial s}, \nonumber \\
\bar{X}_2 f(s,k) &=& \tau s f(s,k) ,
\eea
and find they are in fact particular solution, 
with $R_1=\nu\gamma$ and $R_2 = \frac{\tau}{2\pi}$.
We note their commutator is
\be
[ \bar{X}_1 , \bar{X}_2 ] = \frac{2 \pi i}{\gamma} R_1 R_2 .
\ee
Later we will interprete the solution as gauge bundle on the
dual space of compactification, and above particular solution
is one with constant curvature.

Now we are looking for two independent operators which 
commutes with both $U_1,U_2$, which will be generators
for the gauge field. 
\bea
Z_1 f(s,k) &=& e^{2 \pi i s /q} f(s,k-1), \nonumber \\
Z_2 f(s,k) &=& e^{i \nu k} f(s+ \sigma , k), 
\eea
with $\sigma = \frac{1}{\gamma q}, \nu = - \frac{2\pi a}{q}$ 
where $ap + bq = 1$ and $a,b \in \bf Z$. They satisfy
\be
Z_1 Z_2 = e^{2 \pi i \theta'} Z_2 Z_1 ,
\ee
with 
\be
\theta' = \frac{a \theta + b}{p - q \theta } \label{sl} .
\ee
Thus the homogeneous solution of compactified directions $A_i$,
and $X_a$ for uncompactified directions are thought to be fields
on the dual noncommutative torus with parameter $\theta'$.
Above discussion constitutes rough sketch of Morita equivalence;
U($q$) theory on ${\bf T}_{-\theta}$ is identical to U(1) theory
on ${\bf T}_{\theta'}$, where $\theta$ and $\theta'$ are related
by SL(2,${\bf Z}$) transformation (\ref{sl}). 
For U(1) theory the two generators $Z_1,Z_2$
will be interpreted as
\be
Z^m_1 Z^n_2 \ra e^{i(m \sigma_1 + n \sigma_2 + \pi \theta' mn )} ,
\ee
where $\sigma_i$ are coordinates of the dual noncommutative torus,
satisfying
\be
[\sigma_1 , \sigma_2 ] = - 2\pi i \theta' .
\ee
And the general solution of the compactification condition is
\bea
X_1 &=& \bar{X}_1 + \sum c_{mn} {\cal Z}_{mn}, \nonumber \\
X_2 &=& \bar{X}_2 + \sum d_{mn} {\cal Z}_{mn}, 
\eea
where
\be
{\cal Z}_{mn} = e^{-\pi i m n \theta'} Z^m_1 Z^n_2.
\ee
Since
\bea
\ [\bar{X}_1 , Z_1] &=& -\frac{2 \pi \nu}{q} Z_1 , \nonumber \\  
\ [ \bar{X}_2 , Z_2 ] &=& - \tau \sigma Z_2 ,
\eea
We can identify as
\be
\bar{X}_i \ra i \frac{2\pi R_i}{\gamma q} D_i ,
\ee
with constant curvature
\be
[ D_1 , D_2 ] = \frac{\gamma q^2}{2 \pi i} .
\ee
And the general solutions should be identified as
\bea
X_1 &=&
 i \frac{2 \pi R_1}{\gamma q} D_1 
+ A_1 (\sigma_1, \sigma_2) ,
\nonumber \\
X_2 &=&
 i \frac{2 \pi R_2}{\gamma q} D_2
+ A_2 (\sigma_1, \sigma_2) ,
\eea
where $A_i$ are gauge field defined on a noncommutative torus.
\section{Orientifolds of CDS' solution}
\subsection{Cylinder}
In addition to the toroidal compactification condition
(\ref{comdef}), we impose one more condition to make it
orientifold on cylinder,
\bea
{\cal M} X_1 {\cal M}^{-1} &=& -X^T_1 \nonumber ,\\
{\cal M} X_2 {\cal M}^{-1} &=& +X^T_2 .
\eea
This is Matrix theory realization of the involution giving
rise to cylinder from torus,
\be
(\sigma_1 , \sigma_2 ) \sim ( - \sigma_1 , \sigma_2).
\ee
Considering successive transformations we can find consistency
condition, following \cite{howuwu}.
\bea
U_1 U_2 &=& e^{2 \pi i \theta} U_2 U_1, \\
U^*_1 {\cal M} &=& \epsilon_1 {\cal M} U^{-1}_1 \label{cyl1}, \\
U^*_2 {\cal M} &=& \epsilon_2 {\cal M} U_2, \\
{\cal M} {\cal M}^* &=& \epsilon {\bf 1} \label{cylu2}.
\eea
All new parameters introduced here are complex numbers with
unit magnitude. It is obvious that $\epsilon_2$ can be scaled away by
redefining $U_2$. It turns out that to satisfy the consistency conditions,
we can choose only from $\epsilon_1 = \pm 1$,$\epsilon = \pm 1$. 
It was found that
$\epsilon_1=1$ corresponds to the case that the dual space is cylinder,
while when $\epsilon=-1$ we have Klein bottle instead \cite{howuwu}. 
$\ep$ seems to select
the gauge group on the boundary, when the dual space is cylinder.
We will study this further in the following sections on quantum bundle.

Now that we have found the solutions for $U_1,U_2$ as operators on
functions defined on ${\bf R} \otimes {\bf Z}_q$, our next task here is 
to find ${\cal M}$. Eq.(\ref{cylu2})
means that when $\ep=-1$, ${\cal M}$ is antisymmetric. We know that
antisymmetric matrices of odd dimensionality cannot be unitary.
And Eq.(\ref{cyl1}) amounts to finding unitary transformation
between $U_1$ and $U^T_1$, when $\ep_1=1$. But it turns out that this
cannot be done when ${\cal M}$ is antisymmetric. 
So we have solutions for only three cases.

Let's begin with the case of $(\ep_1,\ep)=(1,1)$.
The solution is
\be
{\cal M} f(s,k) = f(s,-k),
\ee
Under which 
\bea
{\cal M} Z_1 {\cal M}^{-1} &=& Z^T_1, \nonumber \\
{\cal M} Z_2 {\cal M}^{-1} &=& Z^*_2, 
\eea
which means
\be 
{\cal M} ({\cal Z}_{mn}) {\cal M}^{-1} = ({\cal Z}_{m,-n})^T.
\ee
So if we identify the operators $Z_1,Z_2$ as generators of U(1) bundle
on the noncommutative torus, functions which are invariant under the
projection condition should satisfy
\bea
A_1(\sigma_1, \sigma_2) &=& - A_1(\sigma_1 , -\sigma_2) \nonumber \\
A_2(\sigma_1, \sigma_2) &=& + A_2(\sigma_1 , -\sigma_2) 
\label{cylres1}
\eea
So the dual space is cylindrical, as expected.

Now turn to the case of $(\ep_1,\ep)=(-1,1)$.
Looking at Eq.(\ref{cyl1}) and taking determinant, we can
show that $q$ should be even. And since $p$ is prime to $q$,
it is odd. The solution is
\be
{\cal M} f(s,k) = (-1)^k f(s,-k).
\ee
And under that
\bea
{\cal M} Z_1 {\cal M}^{-1} &=& - Z^T_1, \nonumber \\
{\cal M} Z_2 {\cal M}^{-1} &=& Z^*_2 .
\eea
So on the dual noncommutative two torus we have the following
conditions,
\bea
A_1(\sigma_1, \sigma_2) &=& 
- A_1(\sigma_1 + \pi, -\sigma_2), \nonumber \\
A_2(\sigma_1, \sigma_2)  &=& 
+ A_2(\sigma_1 + \pi, -\sigma_2). 
\eea
So we have Klein bottle.

Last solution for $(-1,-1)$. Again $q$ is even, while $p$ odd. 
The solution is
\be
{\cal M} f(s,k) = (-1)^k  f(s,1-k),
\ee
and 
\bea
{\cal M} Z_1 {\cal M}^{-1} &=& - Z^T_1, \nonumber \\
{\cal M} Z_2 {\cal M}^{-1} &=& e^{2 \pi i \frac{a}{q}} Z^*_2. 
\eea
So on the dual noncommutative two torus we have the following
conditions,
\bea
A_1(\sigma_1, \sigma_2) &=& 
- A_1(\sigma_1 + \pi, 2 \pi q/a -\sigma_2), \nonumber \\
A_2(\sigma_1, \sigma_2) &=& 
+ A_2(\sigma_1 + \pi, 2 \pi q/a  -\sigma_2). 
\eea
Above boundary conditions give Klein bottle as well. 
Or the coefficient in front
of $Z^*_2$ here may be thought to be irrelevant, because we can scale it
away redefining $Z_2$. The convenient fundamental region of the
half-torus could be different, but the topology is intact.
\subsection{Klein Bottle}
Now we consider orientifolding on a Klein bottle.
\bea
{\cal M} X_1 {\cal M}^{-1} &=& - X^T_1 , \nonumber \\
{\cal M} X_2 {\cal M}^{-1} &=& + X^T_2 + \pi R_2,
\eea
which are Matrix theory realization of the following involution
making Klein bottle from torus,
\be
(\sigma_1,\sigma_2) \sim (- \sigma_1, \pi + \sigma_2).
\ee
Now we follow the same procedure we used for the case of cylinder.
The consistency consideration gives
\bea
U_1 U_2 &=& e^{2\pi i \theta} U_2 U_1, \\
U^*_1 {\cal M} &=& \ep_1 {\cal M} U^{-1}_1, \\
U^*_2 {\cal M} &=& \ep_2 {\cal M} U_2, \\
{\cal M}^* {\cal M} &=& \ep U_2. 
\eea
This time both $\ep_2,\ep$ can be absorbed into $U_2$, so irrelevant. 
And consistency consideration gives us $\ep^2_1 = e^{-2\pi i \theta}$.
A solution can be found only if $q$ is odd, with $\epsilon_1 = 
(-1)^p e^{-\pi i \theta}, \epsilon_2=\epsilon=1$.
\be
{\cal M} f(s,k) = (-1)^k e^{\frac{\pi i k}{q}} f(s+ 1/2 , -k),
\ee
this transforms the basis 
\bea
{\cal M} Z_1 {\cal M}^{-1} &=& -Z^T_1, \nonumber \\
{\cal M} Z_2 {\cal M}^{-1} &=& Z^*_2.
\eea
It is evident that we have Klein bottle for the dual space. 
This is a rather surprising result, since for U(1) theory on
noncommutative torus, the orientifold projection apparently
could give cylindrical or Klein bottle topology for the dual
space \cite{howuwu}. But when we actually try to find the solution, we have
only one case, which gives SYM theory on Klein bottle.
\subsection{M\"obius strip}
The involution we have to realize in terms of matrices is
\be
(\sigma_1, \sigma_2 ) \sim (\sigma_2, \sigma_1).
\ee
The orientifold condition is
\bea
{\cal M} X_1 {\cal M}^{-1} &=& \frac{R_1}{R_2} X^T_2, \nonumber \\
{\cal M} X_2 {\cal M}^{-1} &=& \frac{R_2}{R_1} X^T_1.
\eea
The consistency condition gives
\bea
U_1 U_2 &=& e^{2 \pi i \theta} U_2 U_1, \\
U^*_1 {\cal M} &=& \epsilon_{1} {\cal M} U_2,  \\
U^*_2 {\cal M} &=& \epsilon_{2} {\cal M} U_2,  \\
{\cal M}^* {\cal M} &=& \epsilon \bf 1.
\eea
We can find one solution which is Fourier transformation operator 
on all the variables,
\be
{\cal M} f(s,k) = \int dt \sum_{l=1}^{q} e^{2 \pi i \gamma st - i \nu k l} f(t,l)
\ee
with $\epsilon_1 = \epsilon_2 = \epsilon = 1$, which gives
\bea
{\cal M} Z_1 {\cal M}^{-1} &=& Z^*_2, \nonumber \\
{\cal M} Z_2 {\cal M}^{-1} &=& Z^*_1.
\eea
It is straightforward to check that the general should satisfy
\bea
A_1(\sigma_1 , \sigma_2 ) &=& A_2(-\sigma_2,-\sigma_1), \nonumber \\
A_2(\sigma_1 , \sigma_2 ) &=& A_2(-\sigma_2,-\sigma_1),
\eea
which defines dual M\"obius strip through boundary condition.
\section{Twisted Quantum Bundle on $\bf T^2$ }
In this section we review the construction of twisted quantum U($q$) 
bundle on noncommutative torus with constant abelian curvature. 
This is studied first in \cite{ho} and generalized later in \cite{bmz}.
Quantum torus ${\bf T}^2_{-\theta}$ is defined in terms of two
noncommuting coordinates, i.e.
$$
[ \sigma_1 , \sigma_2 ] = 2 \pi i \theta .
$$
Using gauge invariance, any connection with constant curvature
can be written as 
\be
D_1 = \partial_1 + i F \sigma_2,  \;\;\;
D_2 = \partial_2 - i F \sigma_1, \label{qconn}
\ee
with field strength 
\be
{\cal F} \equiv i[D_1, D_2] = 2 (F + \pi \theta F^2 ) .
\ee
The connections satisfy the periodic boundary condition up to unitary
transition functions $\Omega_i$.
\bea
D_i(\sigma_1 + 2 \pi, \sigma_2) &=&
\Omega_1 (\sigma_2) D_i(\sigma_1,\sigma_2) \Omega_1^{-1} (\sigma_2),
\nonumber \\
D_i(\sigma_1 , \sigma_2 + 2 \pi) &=&
\Omega_2 (\sigma_1) D_i(\sigma_1,\sigma_2) \Omega_2^{-1} (\sigma_1).
\label{qbundle}
\eea
The solutions for $\Omega_i$ can be found easily,
\bea
\Omega_1 &=& e^{iP \sigma_2} U, \nonumber \\
\Omega_2 &=& e^{-iP \sigma_1} V, 
\eea
where 
\be
P = \frac{2\pi F}{1 + 2 \pi \theta F},
\ee
and $U,V$ are q-dimensional 't Hooft matrices satisfying
\bea
U V = \omega V U \nonumber,
\eea
with $\omega=e^{-2\pi i p/q}$. In this paper we choose
$U_{ij} = \omega^i \delta_{ij}$ and $V_{ij} = \delta_{i,j+1}$.
Without loss of generality we assume $q,p$ are 
relatively prime. 

Due to the requirement of consistency 
the transition functions $\Omega_i$ must satisfy the cocycle condition
\be
\Omega_1 (\sigma_2 + 2 \pi ) \Omega_2 (\sigma_1)
=
\Omega_2 (\sigma_1 + 2 \pi ) \Omega_1 (\sigma_2).
\ee
This imposes the following conditions
\bea
\frac{p}{q} \theta &=& 1 - Q^{-2}, \nonumber \\
Q &=& 1 + 2 \pi \theta F.
\eea
Now we can find the adjoint section of this quantum bundle,
which satisfy (\ref{qbundle})
\bea
Z_1 &=& e^{iQ\sigma_1 /q} V^b, \nonumber \\
Z_2 &=& e^{iQ\sigma_2 /q} U^{-b}, 
\eea
where $a,b$ are integers and satisfy $aq - bp = 1$.
$Z_1,Z_2$ generate the algebra of sections on the adjoint bundle.
They satisfy
$$
Z_1 Z_2 = e^{2\pi i \theta'} Z_2 Z_1 ,
$$
with
\be
\theta' = \frac{a (-\theta) + b}{p (-\theta) + q}.
\ee
The general solution of U($q$) quantum bundle on noncommutative
torus ${\bf T}^2_{-\theta}$ can be written as
\be
A_i (\sigma_1 , \sigma_2 ) =
\sum_{i,j \in {\bf Z}} c_{mn}
{\cal J}_{mn} (\sigma_1, \sigma_2),
\ee
where
\bea
{\cal J}_{mn} (\sigma_1 , \sigma_2 )
&=& e^{-\pi i m n \theta'} Z_1^m (\sigma_1) Z_2^n (\sigma_2),
\nonumber \\
&=& J_{mn} e^{iQ (m\sigma_1 + n\sigma_2 )/q},
\eea
with
\be
J_{mn} = e^{-\pi i  mn b/q} V^{bm} U^{-bn}.
\ee
The duality of SYM on noncommutative torus comes from the fact that
${\cal J}_{mn}$ can be treated as U(1) bundle on 
${\bf T}^2_{\theta'}$ as well as U($q$) bundle on ${\bf T}_{-\theta}$. 
Note that $J_{mn}$ generate U($q$). It is obvious that $\theta$
and $\theta'$ are related by SL(2,${\bf Z}$). Actually it is proved
\cite{riesch} that in general the duality group is SO($d,d|{\bf Z}$) on 
$d$-dimensional noncommutative torus. To be correct what was shown is that two 
noncommutative tori ${\bf T}^d_{\theta}$ and ${\bf T}^d_{\hat{\theta}}$ are
Morita equivalent when $\theta$ and $\hat{\theta}$ belong to the same
orbit of the group SO$(d,d|{\bf Z})$, the 
and equivalence of action functionals \cite{schwarz}
and BPS spectrum of SYM theories on Morita equivalent tori are proved
\cite{ho,bm,konsch}.  For two dimensional case 
SO$(2,2|{\bf Z})=$ SL(2,${\bf Z})\times {\rm SL}(2,{\bf Z}$), 
where one SL$(2,{\bf Z})$ is the ordinary symmetry for 
two dimensional torus, and the other SL$(2,{\bf Z})$ is 
T-duality which involves the 
lightlike direction. In the following two sections we will study
the orientifolding of this quantum twisted bundle on cylinder
and Klein bottle respectively.

\section{Twisted Quantum bundle on Cylinder}
The twisted boundary condition should be
\bea
D_1^T (\sigma_1 , -\sigma_2 ) &=&
- {\cal M} D_1(\sigma_1,\sigma_2) {\cal M}^{-1}, 
\nonumber \\
D_2^T (\sigma_1 , -\sigma_2 ) &=&
+ {\cal M} D_2(\sigma_1,\sigma_2) {\cal M}^{-1} \label{cyl},
\eea
We can introduce coordinate dependence into ${\cal M}$, but
it turns out it that does not give us any genuine physical difference.
Now ${\cal M}$ is a unitary matrix and acts only on the 
gauge group part, and it is straightforward to check that the constant 
curvature connection (\ref{qconn})
satisfies above conditions with chosen sign convention, 
which is consistent with our previous result (\ref{cylres1}). 

Here we must consider additional consistency conditions
which is similar to the cocycle condition. First we act the 
orientifold condition twice and get
\be
{\cal M}^* \; {\cal M} = \pm \; {\bf 1},
\ee
which means ${\cal M}$ is either symmetric or antisymmetric.

Mingled with $\Omega_i$, we also get 
\bea
{\cal M} \; \Omega_1 (\sigma_2) {\cal M}^{-1}
&=&
e^{i \phi_1} \; \Omega^*_1 (-\sigma_2) ,
\nonumber \\
{\cal M} \; \Omega^{-1}_2 (\sigma_1) {\cal M}^{-1} 
&=&
e^{i \phi_2} \; \Omega^*_2 (\sigma_1) ,
\eea
where $\phi_i$ are arbitrary phases.
Coordinate dependence is trivially satisfied, and
the gauge part is essentially
the same with the solutions we found before for orientifolding
of CDS' projective module solution.

As symmetric one, we have
\be
{\cal M} = \pmatrix{1 &&&& \cr &&&& 1 \cr &&& 1  & \cr
&& \cdots && \cr  & 1 &&&}.
\ee 
Then the adjoint section we found before transforms as
\bea
{\cal M} Z_1 (\sigma_1,\sigma_2) {\cal M}^{-1} &=& 
Z_1^T (\sigma_1 , -\sigma_2 ),
\nonumber \\
{\cal M} Z_2 (\sigma_1,\sigma_2) {\cal M}^{-1} &=& 
Z_2^{-1T} (\sigma_1 , -\sigma_2 ).
\eea
Thus we find the solution after orientifolding as 
\be
A_i (\sigma_1 , \sigma_2 ) = \sum_{m,n \in {\bf Z}}
c_{m,n} \left( {\cal J}_{m,n}  + (-1)^i {\cal J}_{m,-n} \right).
\ee
It is important to check what happens on the boundary of the
cylinder, i.e. $\sigma_2 = 0$.
\be
A_i (\sigma_1 , 0 ) = \sum_{m,n \in {\bf Z}}
c_{mn} \left( J_{m,n}  + (-1)^i J_{m,-n} \right) 
e^{iQm\sigma_1 /q} .
\ee
It is known that $J_{m,n}-J_{m,-n}$ generate SO($q$) \cite{kimrey1}. 
$A_1$ is in adjoint and $A_2$ is in symmetric tensor representation.

Next two choices apply only when $q$ is even. First we have
\be
{\cal M} = \pmatrix{1 &&&& \cr &&&& -1 \cr &&& 1  & \cr
&& \cdots && \cr  & -1 &&&}.
\ee 
Then we have
\bea
{\cal M} Z_1 (\sigma_1,\sigma_2) {\cal M}^{-1} &=& 
(-1)^b Z_1^T (\sigma_1 , -\sigma_2 ),
\nonumber \\
{\cal M} Z_2 (\sigma_1,\sigma_2) {\cal M}^{-1} &=& 
Z_2^{-1T} (\sigma_1 , -\sigma_2 ).
\eea
Since $q$ is even, $b$ is always odd, which is obvious from
$aq-bp=1$.
Thus we have the general solution as following.
\be
D_i (\sigma_1 , \sigma_2 ) = \sum_{m,n \in {\bf Z}}
c_{m,n} \left( {\cal J}_{m,n}  + (-1)^i (-1)^m {\cal J}_{m,-n} \right).
\ee
On the boundary
\be
D_i (\sigma_1 , 0 ) = \sum_{m,n \in {\bf Z}}
c_{mn} \left( J_{m,n}  + (-1)^i (-1)^m J_{m,-n} \right) 
e^{iQm\sigma_1 /q} .
\ee
Within our choice of $U,V$ this subset corresponds to SO($q$) \cite{kimrey1}.

Finally we have to try the case when ${\cal M}$ is antisymmetric,
\be
{\cal M} = \pmatrix{ &&&& 1 \cr &&& -1 & \cr && \cdots & & \cr
& 1 &&& \cr   -1 &&&&}.
\ee 
Since we have
\bea
{\cal M} U {\cal M}^{-1} &=& e^{2\pi i p/q} U^{-1},
\nonumber \\
{\cal M} V {\cal M}^{-1} &=& - V^{-1}.
\eea
We get 
\bea
{\cal M} Z_1 (\sigma_1,\sigma_2) {\cal M}^{-1} &=& (-1)^b
Z_1^T (\sigma_1 , -\sigma_2 ),
\nonumber \\
{\cal M} Z_2 (\sigma_1,\sigma_2) {\cal M}^{-1} &=& e^{-2\pi i b p /q}
Z_2^{-1T} (\sigma_1 , -\sigma_2 ).
\eea
And we again make use of the fact that $b$ is odd, and $bp=aq-1$ to get
the general solution
\be
A_i (\sigma_1 , \sigma_2 ) = \sum_{m,n \in {\bf Z}}
c_{mn} \left( {\cal J}_{mn}  + 
(-1)^i (-1)^m e^{2\pi i n/q} {\cal J}_{m,-n} \right).
\ee
On the boundary
\be
A_i (\sigma_1 , 0 ) = \sum_{m,n \in {\bf Z}}
c_{mn} \left( J_{mn}  + (-1)^i (-1)^m e^{2\pi i n/q} J_{m,-n} \right) 
e^{iQm\sigma_1 /q} .
\ee
And with our choice we have USp($q$) on the boundary \cite{kimrey1}.

To summarize, twisted U($q$) bundle on cylinder can have SO($q$) or USp($q$)
on the boundary according to the solutions.
\section{Twisted Quantum bundle on Klein Bottle}
We consider
\bea
D_1^T (\sigma_1 + \pi  , -\sigma_2 ) &=&
- {\cal M} (\sigma_1,\sigma_2) D_1(\sigma_1,\sigma_2) 
{\cal M}^{-1} (\sigma_1,\sigma_2),
\nonumber \\
D_2^T (\sigma_1 + \pi  , -\sigma_2 ) &=&
+ {\cal M} (\sigma_1,\sigma_2) D_2(\sigma_1,\sigma_2) 
{\cal M}^{-1} (\sigma_1,\sigma_2).
\eea
We have to consider the consistency condition. As the first
one we act ${\cal M}$ twice on $D_i$ and identify with $\Omega_2$. 
Then we have
\be
{\cal M} (\sigma_1 , \sigma_2 ) =
e^{iP \sigma_2 /2} {\cal M}_{KB},
\ee
with
\be
{\cal M}_{KB}^* {\cal M}_{KB} = U \label{klein}.
\ee 
where $U$ is the gauge part of $\Omega_2$.
Again if we consider successive transformations with 
$\Omega_i$ and ${\cal M}$, we get 
\bea
{\cal M} \; \Omega_1 (\sigma_2) {\cal M}^{-1}
&=&
e^{i \phi_1} \; \Omega^*_1 (-\sigma_2) ,
\nonumber \\
{\cal M} \; \Omega^{-1}_2 (\sigma_1) {\cal M}^{-1} 
&=&
e^{i \phi_2} \; \Omega^*_2 (\sigma_1 + \pi), 
\eea
where $\phi_i$ are arbitrary phases.
When $q$ is odd we can easily find the solution to (\ref{klein}).
It is done as follows. $U,V$ are related by unitary transformation,
so assume $K U K^{-1} = V$, then we can easily check ${\cal M}_{KB}=
K^T V^{\frac{q+1}{2}} K$ satisfies (\ref{klein}). 
It turns out that when $q$ is even one
cannot find any solution, which is consistent with the fact that 
with the conventional choice of $U,V$
the determinant of $U$ is $-1$ when $q$ is even, while from the left
hand side it should be always positive. Our choice in this paper is
$U=U^T$ and $V^T=V^{-1}$ so the solution for ${\cal M}$ is found to be 
\be
{\cal M}_{KB} =
\pmatrix{1 &&&& \cr &&&& \eta^{q-1} \cr
&&& \cdots & \cr && \eta^2 && \cr 
& \eta },
\ee
where $\eta=\omega^\frac{q+1}{2}$.
Then we have
\bea
{\cal M}_{KB} U {\cal M}^{-1}_{KB} &=& U^{-1}, \nonumber \\
{\cal M}_{KB} V {\cal M}^{-1}_{KB} &=& \eta V^{-1} .
\eea
And after some calculation the transformation for $Z_i$
are simplified as
\bea
{\cal M} (\sigma_1 , \sigma_2 ) 
Z_1 (\sigma_1 , \sigma_2 )
{\cal M}^{-1} (\sigma_1 , \sigma_2 )
&=&
- Z_1^T (\sigma_1 + \pi , -\sigma_2 ),
\nonumber \\
{\cal M} (\sigma_1 , \sigma_2 ) 
Z_2 (\sigma_1 , \sigma_2 )
{\cal M}^{-1} (\sigma_1 , \sigma_2 )
&=&
 Z_2^{-1T} (\sigma_1 + \pi , -\sigma_2 ).
\eea
Thus the general solution should be written as
\be
A_i (\sigma_1 , \sigma_2 ) = \sum_{m,n \in {\bf Z}}
c_{mn} \left( {\cal J}_{mn} + 
(-1)^i (-1)^m {\cal J}_{m,-n} \right).
\ee
\section{Discussions}
In this paper we studied aspects of 
Matrix theory orientifolds on noncommutative torus.
It was important to note that Matrix compactifications can 
allow ambiguity in determining the dual space, but in this paper
concrete solutions may not exist in some cases. 
For example when compactified on Klein bottle, U(1) theory on Klein bottle
can be related to U($q$) with odd $q$ only. Here we used the simplest
projective module solution presented by Connes, Douglas and 
Schwarz \cite{CDS} to investigate the dual space of Matrix theory 
orientifold compactifications. 
Obviously we could extend to more general cases. For example if we consider
functions $f(s_1,s_2,k)$ instead, we surely have noncommutative four-torus.
And we can also introduce more than one gauge indices, e.g. if we consider 
$f(s,k_1,k_2)$ where $k_1 \in {\bf Z}_{q_1}, k_2 \in {\bf Z}_{q_2}$,
it turns out we have Morita equivalence between U(1) and U($q$), where
$q$ is the least common multiple of $q_1,q_2$. The orientifold
operator ${\cal M}$ can act on either of the gauge indices, and there is 
a possibility to obtain Morita equivalent pairs which were excluded by 
our analysis in this paper. Or if our result persists even with more general
projective module solutions there should be physical reason for that.
This is an open problem at this stage, and we hope to report in due time.

Now that we have studied orientifold compactification on ${\bf T}^2$,
it should be very 
interesting to study higher dimensional cases following the procedure
presented here, for example ALE spaces ${\bf C}^2/{\bf Z}_n$ \cite{RW}. We
expect to be able to interpolate the topology of dual space again, 
but when we actually find out the solution some might be excluded
as was the case with lower dimensional examples studied here.
We also suggest the study of heterotic Matrix string theory, which
is Matrix theory compactified on cylinder 
${\bf S}^1 \times {\bf S}^1/{\bf Z}_2$ \cite{lowe,horava,kabrey}, 
with noncommutativity in more detail. To cancel the gauge anomaly
on the boundaries of cylinder we have to introduce fermion fields
which correspond to the D8-branes in type IIA string theory. It is known 
that when the D8-branes are displaced from the boundary we have to introduce 
Chern-Simons term to cancel the gauge anomaly \cite{horava}\cite{kabrey}.
Of course the study of Chern-Simons term on noncommutative torus 
should be very interesting on its own. These issues are under investigation
and we will report somewhere else \cite{hkln}.

\end{document}